# Deep Plug-and-Play HIO Approach for Phase Retrieval

ÇAĞATAY IŞIL,[1] AND FIGEN S. OKTEM,[2,*]

[1] *Dept. of Electrical and Computer Eng., UCLA, Los Angeles, 90095, CA, USA*
[2] *Dept. of Electrical and Electronics Eng., METU, Ankara, 06800, Turkey*
[**] *The results of this manuscript were obtained in 2019-2020 [1, 2].*
[*] *figeno@metu.edu.tr*

**Abstract:** In the phase retrieval problem, the aim is the recovery of an unknown image from intensity-only measurements such as Fourier intensity. Although there are several solution approaches, solving this problem is challenging due to its nonlinear and ill-posed nature. Recently, learning-based approaches have emerged as powerful alternatives to the analytical methods for several inverse problems. In the context of phase retrieval, a novel plug-and-play approach that exploits learning-based prior and efficient update steps has been presented at the Computational Optical Sensing and Imaging topical meeting, with demonstrated state-of-the-art performance. The key idea was to incorporate learning-based prior to the Gerchberg-Saxton type algorithms through plug-and-play regularization. In this paper, we present the mathematical development of the method including the derivation of its analytical update steps based on half-quadratic splitting and comparatively evaluate its performance through extensive simulations on a large test dataset. The results show the effectiveness of the method in terms of both image quality, computational efficiency, and robustness to initialization and noise.

## 1. Introduction

Phase retrieval deals with the recovery of an unknown image from intensity-only measurements. This nonlinear inverse problem arises in a variety of applications such as microscopy [3, 4], crystallography [5], optical imaging [6, 7], and astronomy [8]. In its most commonly encountered form known as Fourier phase retrieval, the available measurements are Fourier intensities. Due to its nonlinear and ill-posed nature, solving the phase retrieval problem, particularly Fourier phase retrieval, is challenging even though a unique solution can be almost always guaranteed in various practical scenarios of interest [9].

Although several solution approaches exist for the phase retrieval problems, each of them has its own drawbacks. Classical approaches are alternating projection-based methods such as the popular Gerchberg-Saxton (GS), error-reduction and hybrid input-output (HIO) algorithms, and their variants [10, 11]. Such projection-based methods are widely used due to their computational efficiency, simple implementation and flexibility to be used for different phase retrieval problems. These methods jointly utilize available intensity measurements and information known a priori, and alternate between space and measurement domains to impose these constraints through projections [10–13]. Alternating projection methods can empirically converge to a reasonably good solution in various applications. However, since regularization is not exploited in their formulation and intensity measurement constraint is non-convex, the reconstructions obtained with these methods often contain errors and artifacts due to noise amplification and being stuck in local minima.

On the other hand, regularization based reconstruction methods are shown to provide better reconstruction than alternating-projection based methods for various inverse problems in imaging. The conventional approach to regularization is to incorporate the prior statistical knowledge of the unknown image in a maximum posterior (MAP) estimation framework. The resulting optimization problem involves a data fidelity term and an analytical regularization term corresponding to the assumed prior knowledge. For phase retrieval, the most prominently used choice is the sparse signal prior such as based on total variation or dictionary learning [14–16]. Instead recent approaches to regularization rely on utilizing highly developed deep image denoisers in place of analytical priors, yielding significant performance improvement in various inverse problems [17], including phase retrieval [18].

One such approach is plug-and-play (PnP) regularization [19–22], which splits the regularized inverse problem into smaller subproblems using variable-splitting methods, such as half quadratic splitting (HQS) [23–25] and alternating direction method of multipliers (ADMM) [19, 26]. The subproblem containing the regularization term then becomes a denoising problem, and is solved with any independently-developed denoising method. Recently deep learning-based denoisers are the prevailing choice as PnP priors due to their state-of-the-art denoising performance. PnP regularization have been applied to several inverse problems [19, 25, 27–33], including Fourier phase retrieval [2, 28] and phase retrieval from coded diffraction patterns (CDPs) [34–38]. *Regularization by denoising (RED)* [39] is another approach that enables the use of an arbitrary denoiser as a regularizer. Different than PnP regularization, RED uses a denoiser to explicitly define the regularization term. RED approach has also been applied to several inverse problems in imaging, including the Fourier phase retrieval problem [18] and phase retrieval from CDPs [18, 40].

Deep prior-based phase retrieval methods typically use few steps of generic optimization methods to obtain an inexact solution for the subproblem containing the data fidelity term. This often makes these methods computationally intensive. Moreover, the performance is often demonstrated using only few test images. There are also other learning-based phase retrieval approaches that purely rely on neural networks to perform inversion or to iteratively refine a rough estimate [41–43]. Such approaches generally have limited reconstruction capability even for simpler (linear) inverse problems than phase retrieval. Recently diffusion-based models, a type of generative models, have also gained popularity for solving inverse problems, with some efforts in phase retrieval [44–47]. Their ability to generate high-quality images from complex data sets has made them a valuable tool in the field. However, due to their generative nature, resulting reconstructions often contain generative artifacts in the form of non-existing realistic features. Moreover, their performance has not been evaluated comprehensively under different noisy measurement settings and through comparison with the state-of-the-art methods using quantitative distortion metrics.

A novel plug-and-play approach that exploits learning-based priors and efficient update steps has been presented at the Computational Optical Sensing and Imaging (COSI) topical meeting [2], which enabled state-of-the-art performance. The key idea here was to incorporate learning-based prior to the hybrid input-output method (HIO) through plug-and-play regularization and hence it is named as *Plug-and-Play HIO*. In this paper, we present the mathematical development of this method including the derivation of its analytical update steps based on half-quadratic splitting and comparatively evaluate its performance through extensive simulations on a large test dataset.

Plug-and-Play HIO utilizes half-quadratic splitting (HQS) to obtain efficient analytical update steps combined with learning-based denoising. Using HQS, the regularized inverse problem is first decoupled into two sub-problems that separately contain the data fidelity and prior terms to utilize PnP regularization. The sub-problem containing the prior term corresponds to a denoising problem and is simply solved with a deep learning-based denoiser. The second sub-problem containing the data fidelity term is solved analytically with efficient measurement and image update steps. The measurement update involves convex combination of the actual measurement and the estimated measurement based on the current guess, with less contribution coming from the actual measurement as the iterations proceed. The image update step performs projections to impose this updated measurement along with the available space domain constraints. By successively repeating the denoising step, measurement update step, and projection steps, the final reconstruction is obtained.

The plug-and-play phase retrieval approach presented in this paper can be viewed as a regularized version of GS-type algorithms with learning-based prior. It demonstrates the strengths of both the classical alternating projection methods and deep learning-based regularization methods. Any denoiser algorithm and GS-type method can be easily exploited within the approach. As a result, the approach is flexible to be used with different measurements including Fourier magnitude and CDP measurements. Using a large test dataset, we illustrate the state-of-the-art performance of the method for the Fourier phase retrieval problem in terms of both computational efficiency, image quality, and robustness to different initialization and noise levels. Early results have also been presented in the COSI 2020 meeting [2].

Contributions of this work can be summarized as follows:

- Development of an efficient model-based phase retrieval method with learning-based denoiser prior that involves analytical image and measurement update steps,

- Comprehensive experiments on a large test dataset to analyze the performance in terms of image quality, computational efficiency, and robustness to different initialization and noise levels.

The paper is organized as follows. Phase retrieval problem is introduced in Section 2 and the plug-and-play phase retrieval approach is presented in Section 3. The performance of the approach is evaluated in Section 4 through a comparative analysis. In Section 5, we provide our final remarks and conclusions.

## 2. Phase Retrieval Problem

In the 2D discrete phase retrieval problem, the goal is to recover an image **x** from intensity-only measurements **y** that can be expressed as

$$\mathbf{y}^2 = |\mathbf{Ax}|^2 + \mathbf{w}, \qquad \mathbf{w} \sim N(\mathbf{0}, \alpha^2 \operatorname{diag}(|\mathbf{Ax}|^2)) \tag{1}$$

where **A** is the $M^2 \times N^2$ system matrix, $\mathbf{x} \in \mathbb{R}^{N^2}$ or $\mathbb{C}^{N^2}$ denotes the vector form of the $N \times N$ unknown image, and $\mathbf{y}^2 \in \mathbb{R}^{M^2}$ represents the vector form of the $M \times M$ noisy intensity measurements, where each vector is obtained using lexicographic ordering. Moreover, $\mathbf{w} \in \mathbb{R}^{M^2}$ denotes the additive Gaussian noise, where $\alpha$ is a scaling parameter that controls the signal-to-noise ratio (SNR) and $\operatorname{diag}(\lambda)$ is a diagonal matrix whose diagonal entries are given by the vector $\lambda$. The measurement noise can also be assumed to have Poisson distribution, and here its normal approximation [18] is used.

An important special case is Fourier phase retrieval problem, for which $\mathbf{A} = \mathbf{F}$ is the 2D DFT matrix. In this most commonly encountered and challenging version, the goal is to recover an image from its Fourier magnitude measurement. Based on the available prior information, the unknown image **x** can be assumed to satify some space domain constraints such as finite support, real-valuedness, and non-negativity, and for uniqueness guarantee, in this case, $M \times M$-point oversampled DFT with $M \geq 2N - 1$ is required [9]. Here we choose $M$ as $2N$ for simplicity.

Alternating projection (AP) algorithms are prevalently used for the solution of the phase retrieval problems due to their flexibility and simple implementation. To recover the missing phase, they iteratively perform projections to exploit the available intensity measurements and prior information. Gerchberg-Saxton (GS) algorithm [12] is one of the earliest alternating projection algorithms, which enforces intensity measurement constraints in the space and Fourier domains to reconstruct a complex-valued image. Error reduction (ER) method is a generalization of the GS algorithm that performs projections to enforce not only the measurement constraints but also any space-domain constraint that is available (such as support and real-valuedness) [13]. Although the error in the recovered intensities is monotonically non-increasing in ER algorithms, recovery of the true solution can not be guaranteed [10]. Due to the intensity (measurement) constraints, these algorithms involve projections onto nonconvex sets and hence can converge to a local minimum.

When space-domain constraints (such as support and real-valuedness) are available in addition to the intensity measurements, the most widely used AP approach is the HIO method [11], which is a relaxed version of the ER method. Similar to the ER, HIO algorithm performs projections for the intensity measurement and space domain constraints. However, different than ER, it does not require the iterates to comply with the space-domain constraints exactly, instead it updates the iterates to finally arrive at a solution that satisfies these constraints. Although the performance of HIO-type methods depends on the initialization and their convergence behavior cannot be analyzed completely, they yet converges to a reasonable solution empirically in various phase retrieval problems [11]. Moreover, empirically they are shown to converge to the global solution when the measurements are noise-free. Nevertheless their reconstructions from noisy measurements suffer from errors and artifacts due to noise amplification and being stuck in local minima.

Alternating projection (minimization)-based phase-retrieval methods are initially developed for Fourier measurements as it arises in astronomical imaging and various imaging applications involving Fraunhofer diffraction such as crystallography. Later, these methods are also extended to general phase retrieval problems [48–50] to handle arbitrary intensity measurements with non-invertible system matrix $\mathbf{A}$. In this general case, the pseudoinverse $\mathbf{A}^\dagger$ is used (instead of matrix inverse/inverse DFT) to go back from the measurement space to the object space. Equivalently, the linear system $\mathbf{A}\mathbf{x} = \mathbf{b}$ can be solved in the least-squares sense with an efficient solver such as conjugate-gradient method (where $\mathbf{b}$ is formed by applying the current phase estimate to the available measurement).

## 3. Plug-and-Play Phase Retrieval Approach via Half Quadratic Splitting

The developed plug-and-play approach is a model-based phase retrieval method that utilizes half-quadratic splitting (HQS) [23] to obtain efficient analytical update steps with learning-based denoising. Below we present the development of the method.

A general formulation of the phase retrieval problem with regularization can be expressed as follows:

$$\hat{\mathbf{x}} = \underset{\mathbf{x}}{\operatorname{argmin}} \frac{1}{2} \||\mathbf{y} - |\mathbf{A}\mathbf{x}|\|^2 + \lambda R(\mathbf{x}) \tag{2}$$

Here $R(\mathbf{x})$ is the regularization function related to the prior information, and $\lambda$ denotes the regularization parameter. This optimization problem is split into sub-problems using HQS. By defining an auxiliary variable $\mathbf{z}$, the problem in Eqn. 2 is first reformulated as follows:

$$\hat{\mathbf{x}} = \underset{\mathbf{x}}{\operatorname{argmin}} \frac{1}{2} \||\mathbf{y} - |\mathbf{A}\mathbf{x}|\|^2 + \lambda R(\mathbf{z}) \quad \text{s.t.} \quad \mathbf{z} = \mathbf{x} \tag{3}$$

This constrained optimization problem is then converted into the following unconstrained form:

$$\hat{\mathbf{x}}, \hat{\mathbf{z}} = \underset{\mathbf{x}, \mathbf{z}}{\operatorname{argmin}} \frac{1}{2} \||\mathbf{y} - |\mathbf{A}\mathbf{x}|\|^2 + \lambda R(\mathbf{z}) + \frac{\mu}{2} \|\mathbf{x} - \mathbf{z}\|^2 \tag{4}$$

where the penalty parameter $\mu$ is non-decreasing. We can solve this lifted problem by performing alternating minimization over $\mathbf{x}$ and $\mathbf{z}$ as follows:

$$\mathbf{x}_{k+1} = \underset{\mathbf{x}}{\operatorname{argmin}} \||\mathbf{y} - |\mathbf{A}\mathbf{x}|\|^2 + \mu_k \|\mathbf{x} - \mathbf{z}_k\|^2 \tag{5a}$$

$$\mathbf{z}_{k+1} = \underset{\mathbf{z}}{\operatorname{argmin}} \frac{\mu_k}{2} \|\mathbf{z} - \mathbf{x}_{k+1}\|^2 + \lambda R(\mathbf{z}) \tag{5b}$$

As seen from above, the regularization and data fidelity terms are decoupled into two simpler sub-problems as desired.

The sub-problem in Eqn. 5b is a denoising problem and can be solved using any Gaussian denoiser $D(\mathbf{x}_{k+1}, \sigma_k)$ where the noise standard deviation $\sigma_k = \sqrt{\lambda/\mu_k}$ should be a non-increasing parameter since $\mu_k$ is non-decreasing. The input and output of the denoiser is $\mathbf{x}_{k+1}$ and $\mathbf{z}_{k+1}$ respectively.

To find the solution of the sub-problem in Eqn. 5a, sub-differential of the objective function at $\mathbf{x}$ should be obtained. For simplicity, here we present the derivation for a real-valued $\mathbf{x}$ and invertible $\mathbf{A}$. The general form of the solution that allow for other cases including complex-valued $\mathbf{x}$ and non-invertible $\mathbf{A}$ is provided afterwards. For the case with real-valued $\mathbf{x}$ and invertible $\mathbf{A}$, following necessary condition is obtained for $\mathbf{x}$:

$$-2\mathcal{R}e\left\{\mathbf{A}^{-1}\left(\mathbf{y} \odot \frac{\mathbf{A}\mathbf{x}}{|\mathbf{A}\mathbf{x}|}\right) - \mathbf{x}\right\} + 2\mu_k(\mathbf{x} - \mathbf{z}_k) = 0 \tag{6}$$

where $\odot$ denotes elementwise multiplication and $\mathcal{R}e\{.\}$ denotes the operator for taking the real-part. This expression suggests a fixed-point iteration, and remarkably initialization of this fixed-point iteration with $\mathbf{z}_k$ provides the following solution that satisfies the above necessary condition for optimality:

$$\mathbf{x}^* = \eta_k \mathcal{R}e\left\{\mathbf{A}^{-1}\left(\mathbf{y} \odot \frac{\mathbf{A}\mathbf{z_k}}{|\mathbf{A}\mathbf{z_k}|}\right)\right\} + (1 - \eta_k)\mathbf{z}_k \tag{7}$$

where $\eta_k = 1/(1 + \mu_k)$. By re-writing this expression, we obtain the measurement and image update steps in our method as follows:

$$\tilde{\mathbf{y}}_k = \eta_k \mathbf{y} + (1 - \eta_k)|\mathbf{A}\mathbf{z}_k| \tag{8}$$

$$\mathbf{x}_{k+1} = \mathcal{R}e\left\{\mathbf{A}^{-1}\left(\tilde{\mathbf{y}}_k \odot \frac{\mathbf{A}\mathbf{z}_k}{|\mathbf{A}\mathbf{z}_k|}\right)\right\} \tag{9}$$

By combining these analytical update steps with the denoising step, the iterations in the developed alternating-minimization approach can be expressed as follows when we have real-valued $\mathbf{x}$ and invertible $\mathbf{A}$:

$$\tilde{\mathbf{y}}_k = \eta_k \mathbf{y} + (1 - \eta_k)|\mathbf{A}\mathbf{z}_k| \tag{10}$$

$$\mathbf{x}_{k+1} = \mathcal{R}e\left\{\mathbf{A}^{-1}\left(\tilde{\mathbf{y}}_k \odot \frac{\mathbf{A}\mathbf{z}_k}{|\mathbf{A}\mathbf{z}_k|}\right)\right\} \tag{11}$$

$$\mathbf{z}_{k+1} = D(\mathbf{x}_{k+1}, \sigma_k) \tag{12}$$

with the parameter $\eta_k$ non-increasing in the range $[0, 1]$ since $\mu_k$ is non-decreasing and non-negative.

Although the derivation above is presented for the simpler case of real-valued $\mathbf{x}$ and invertible $\mathbf{A}$, the approach can be easily generalized for the other cases. Note that the step in Eqn. 11 corresponds to an error-reduction step which performs a projection that applies the current phase estimate to the updated measurement, followed by a second projection that enforces the assumed real-valuedness constraint. To allow for complex-valued $\mathbf{x}$, this second projection can be simply omitted. For the general case with any space-domain constraints (such as support, real-valuedness, nonnegativity, etc.), one can simply replace the operator $\mathcal{R}e\{.\}$ with the respective projection operator $\mathcal{P}_S\{.\}$ for those constraints. Moreover, similar to the well-known extensions of error-reduction algorithm to the general phase retrieval problems [48–50], to handle arbitrary measurements with non-invertible $\mathbf{A}$, the matrix inverse $\mathbf{A}^{-1}$ in Eqn. 11 can be replaced with the pseudoinverse $\mathbf{A}^\dagger$ or equivalently $\mathbf{x}_{k+1}$ can be set to the solution of the linear system $\mathbf{A}\mathbf{x} = \mathbf{b}$ in the least-squares sense where $\mathbf{b}$ is formed by applying the current phase estimate to the updated measurement, i.e. $\mathbf{b} = \tilde{\mathbf{y}}_k \odot \frac{\mathbf{A}\mathbf{z}_k}{|\mathbf{A}\mathbf{z}_k|}$.

The overall algorithm named as *Plug-and-Play Phase Retrieval (PR)* is summarized in Algorithm 1 in its most general form. Note that the algorithm consists of the measurement update step, followed by alternating minimizations/projections to impose the available measurement/space-domain constraints, and then the denoising step. Analytical update steps involve a measurement update based on the estimated measurement for the current guess, followed by the image update step based on alternating projections for the measurement- and space-domain constraints. As given in the 2nd step of Algorithm 1, the measurement is updated at each iteration $k$ as the convex combination of the available measurement $\mathbf{y}$ and the estimated measurement $|\mathbf{Az}_k|$ for the current estimate $\mathbf{z}_k$. Since the parameter $\eta_k$ is non-increasing in the range $[0, 1]$, updated measurement is affected from $|\mathbf{Az}_k|$ more than the actual measurement $\mathbf{y}$ as the iterations proceed. Moreover, the steps 3-6 correspond to the image update that performs projections on the current estimate $\mathbf{z}_k$ similar to the conventional error-reduction (ER) algorithms. These projections apply the current phase estimate to the updated measurement $\tilde{\mathbf{y}}_k$ and then impose the additional constraints in the space domain, if available. Lastly, in the 7th step, this intermediate result is enhanced by a deep denoiser. An appropriate pre-trained denoiser, processing either a real-valued or complex-valued image, is chosen for this purpose depending on the type of the unknown image.

---

**Algorithm 1** Plug-and-Play PR

---

**Input:** $\mathbf{y}, \mathbf{z}_0, T$, non-increasing $\eta_k \in [0, 1]$, non-increasing $\sigma_k \in \mathbb{R}^+$
1: **for** $k = 0$ to $T - 1$ **do**
2: $\quad \tilde{\mathbf{y}}_k \leftarrow \eta_k \mathbf{y} + (1 - \eta_k)|\mathbf{Az}_k|$
3: $\quad \mathbf{x}_{k+1} \leftarrow \mathbf{A}^\dagger \left( \tilde{\mathbf{y}}_k \odot \frac{\mathbf{Az}_k}{|\mathbf{Az}_k|} \right)$
4: $\quad$ **if** space-domain constraints are available **then**
5: $\quad\quad \mathbf{x}_{k+1} \leftarrow \mathcal{P}_\mathcal{S}\{\mathbf{x}_{k+1}\}$
6: $\quad$ **end if**
7: $\quad \mathbf{z}_{k+1} \leftarrow D(\mathbf{x}_{k+1}, \sigma_k)$
8: **end for**

---

When space-domain constraints are available, it is well-known that HIO-type methods have better capability than ER to avoid local minima and to reach a solution that satisfies the nonconvex measurement constraint. For this reason, we also consider a relaxed version of the developed method where the error-reduction updates in steps 3-6 of Algorithm 1 is replaced with a few HIO iterations as given in steps 3-12 in Algorithm 2. This version of the developed method, which is summarized in Algorithm 2, is named as *Plug-and-Play HIO*.

This second method can be viewed as a regularized HIO method that exploits a learning-based prior. In Algorithm 2, the inner HIO iterations use the current estimate $\mathbf{z}_k$ as initialization, and the updated $\tilde{\mathbf{y}}_k$ as the measurement. A learning-based denoiser is then exploited to denoise the final HIO result $\mathbf{v}_L$. These inner and outer iterations continue for a fixed number of iterations $L$ and $T$ respectively. As before, $\sigma_k$ and $\eta_k$ are non-increasing parameters of the denoising and measurement update steps.

As a final remark, we note that for the recovery of complex-valued images, two different approaches can be used for the denoising step. The first involves applying a denoiser separately to the amplitude and phase components, or to the real and imaginary parts of the complex-valued image. This approach uses conventional real-valued DNNs as denoiser thanks to the real-valued nature of these components. Moreover, if these components have characteristics similar to the natural images, the pre-trained denoiser used in this paper can simply be applied. The second approach is to use complex-valued neural networks [51] trained for denoising complex-valued images. While conventional denoisers typically employ real-valued DNNs, their extensions to complex-valued data are also used in different applications. By using either of these two denoising strategies, our method can accommodate the reconstruction of complex-valued images in a wide range of phase retrieval applications.

**Algorithm 2** Plug-and-Play HIO
___
**Input:** $\mathbf{y}$, $\mathbf{z}_0$, $T$, $L$, $\beta$, non-increasing $\eta_k \in [0, 1]$, non-increasing $\sigma_k \in \mathbb{R}^+$
1: **for** $k = 0$ to $T - 1$ **do**
2:      $\tilde{\mathbf{y}}_k \leftarrow \eta_k \mathbf{y} + (1 - \eta_k)|\mathbf{A}\mathbf{z}_k|$
3:      $\mathbf{v}_0 \leftarrow \mathbf{z}_k$
4:      **for** $i = 0$ to $L - 1$ **do**
5:         $\mathbf{v}_{i+1} \leftarrow \mathbf{A}^\dagger \left( \tilde{\mathbf{y}}_k \odot \frac{\mathbf{A}\mathbf{v}_i}{|\mathbf{A}\mathbf{v}_i|} \right)$
6:         **if** space-domain constraints are available **then**
7:            $\gamma \leftarrow$ set of indices of $\mathbf{v}_{i+1}$ that violates the space-domain constraints (support, real-valuedness, non-negativity, etc.)
8:            $\mathbf{v}'_{i+1}[n] = \begin{cases} \mathbf{v}_{i+1}[n] & \text{for } n \notin \gamma \\ \mathbf{v}_i[n] - \beta \mathbf{v}_{i+1}[n] & \text{for } n \in \gamma \end{cases}$
9:            $\mathbf{v}_{i+1} \leftarrow \mathbf{v}'_{i+1}$
10:         **end if**
11:      **end for**
12:      $\mathbf{x}_{k+1} \leftarrow \mathbf{v}_L$
13:      $\mathbf{z}_{k+1} \leftarrow D(\mathbf{x}_{k+1}, \sigma_k)$
14: **end for**
___

## 4. Numerical Results

To show the effectiveness of the developed methods in terms of noise tolerance, computational efficiency, and image generality, we consider the challenging Fourier phase retrieval problem where real-valued images are reconstructed from their Fourier measurements. This problem is encountered in various applications including astronomical imaging [52], laser-illuminated imaging [53], and imaging through random scattering media [54]. In our analysis, we use a large test dataset from [43] that contains 230 *natural* images. This consists of 200 test images of BSD dataset, 24 images of Kodak dataset [55], and 6 natural images used for testing PrDeep [18]. To also evaluate the performance of the approach for images other than natural, we use six images taken by telescopes and scanning electron microscopes, which correspond to the *unnatural* image dataset of PrDeep [18]. We refer these 6 images as *other* in this work to avoid any misconception. All images are of size $256 \times 256$ and have pixel values between 0 and 255.

The noisy Fourier intensity measurements are generated using Eqn. 1 with $\alpha = 2, 3, 4$, which is equivalent to an average SNR of 33.39, 31.67, 30.44 dB respectively (where SNR is defined as $10 \log(\left\||\mathbf{Fx}|^2\right\|_2 / \|\mathbf{w}\|_2)$). These measurements are used with a particular initialization procedure proposed in [18]. In this procedure, the HIO method is first run with 50 different random initializations for 50 iterations. Then, the reconstruction $\hat{\mathbf{x}}$ with the lowest residual ($\||\mathbf{F}\hat{\mathbf{x}}| - \mathbf{y}\|_2$) is processed with additional 1000 HIO iterations. The final result is considered as the reconstruction of the HIO method in our comparisons. This is also used for the initialization of all the other algorithms including the developed plug-and-play methods as well as all the compared algorithms which are prDeep [18] and iterative HIO-DNN method [43].

To enable a fair comparison, we used the respective authors' implementations for the compared learning-based methods. Iterative DNN-HIO method is not customized since in [43] its performance was tested on the same test dataset and with the same initialization procedure as here. PrDeep also uses the same initialization procedure in [18]. PrDeep's parameter $\lambda$, which determines the amount of regularization, was set to the sample variance of the noise, as required in [18] when dealing with Fourier measurements.

For comparisons, peak signal-to-noise ratio (PSNR) and structural similarity index (SSIM) [56] are used as quantitative metrics to comparatively evaluate the performance of all methods. Sample reconstructions are also provided for visual evaluation. All computations are performed using

MATLAB with MatConvNet toolbox [57] and Nvidia Quadro P5000 GPU.

As deep denoiser prior, a pre-trained CNN-based denoiser developed in [25] is used in the plug-and-play PR and plug-and-play HIO methods. This denoiser consists of 25 DNNs trained for different Gaussian noise levels with standard deviations chosen from the interval [1, 50] with a step size of 2. These DNNs involves dilation of convolution layers, residual learning to improve image denoising performance, and batch normalization [25, 58]. More details of the network architecture can be found in [25].

In the developed methods, there are parameters that should be picked such as the number of inner iterations $L$, outer iterations $T$, and the HIO update parameter $\beta$. The parameters $L$ and $T$ are respectively chosen as 5 and 200 since it is observed that they are generally sufficient for the convergence of Plug & Play HIO. The HIO parameter $\beta$ is selected as its common value 0.9. As discussed before, $\sigma_k$ and $\eta_k$ are non-increasing parameters used in the denoising and image update steps with $\eta_k$ taking values in the range [0, 1]. We choose $\sigma_k$ as logarithmically decreasing from a pre-determined maximum value $\sigma_{max}$ to a minimum value $\sigma_{min}$ throughout iterations. The maximum and minimum values, $\sigma_{max}$ and $\sigma_{min}$, that determine the range of the parameter $\sigma_k$, are optimized through a numerical search and set to 40 and 5, respectively. For easier parameter tweaking, the normalized values $\sigma_k/\sigma_{max}$ are used for the parameter $\eta_k$.

The average reconstruction performance of different methods are evaluated for 236 different test images using 5 Monte Carlo runs, and are given in Table 1 for different Poisson noise levels with $\alpha = 2, 3, 4$. The average runtime of each method is also listed in this table. As seen from the table, plug-and-play HIO shows the best overall reconstruction quality for natural images in terms of both SSIM and PSNR, with little additional runtime compared to the conventional HIO. In fact, although the Plug-and-Play HIO is slightly slower than the iterative DNN-HIO method, it is approximately 2-fold faster than PrDeep in addition to providing the best reconstruction performance among the compared methods.

Table 1. **The average reconstruction performance and runtime for** 236 **test images in** 5 **Monte Carlo runs**

| $\alpha = 2$ | Avg. PSNR (dB) | | | Avg. SSIM | | | Runtime |
|---|---|---|---|---|---|---|---|
| (Avg. SNR: 33.39 dB) | Overall | Natural | Other | Overall | Natural | Other | (sec.) |
| HIO | 20.90 ± 0.10 | 20.87 ± 0.10 | 21.74 ± 0.81 | 0.429 ± 0.003 | 0.428 ± 0.003 | 0.455 ± 0.027 | **162** |
| PrDeep [18] | 25.78 ± 0.29 | 25.83 ± 0.29 | 23.92 ± 1.77 | 0.663 ± 0.008 | 0.666 ± 0.008 | 0.565 ± 0.041 | 475 |
| Iterative DNN-HIO [43] | 25.63 ± 0.19 | 23.63 ± 0.18 | 25.47 ± 1.37 | 0.737 ± 0.006 | 0.738 ± 0.006 | **0.691 ± 0.037** | 173 |
| Plug-and-Play PR | 22.10 ± 0.10 | 22.07 ± 0.10 | 23.06 ± 1.26 | 0.597 ± 0.005 | 0.597 ± 0.005 | 0.596 ± 0.023 | 201 |
| Plug-and-Play HIO | **26.35 ± 0.24** | **26.36 ± 0.24** | **26.17 ± 2.22** | **0.751 ± 0.008** | **0.752 ± 0.008** | 0.691 ± 0.041 | 247 |
| $\alpha = 3$ | Avg. PSNR (dB) | | | Avg. SSIM | | | Runtime |
| (Avg. SNR: 31.67 dB) | Overall | Natural | Other | Overall | Natural | Other | (sec.) |
| HIO | 19.79 ± 0.14 | 19.76 ± 0.15 | 20.99 ± 0.81 | 0.356 ± 0.003 | 0.355 ± 0.003 | 0.412 ± 0.030 | **163** |
| PrDeep [18] | 24.22 ± 0.35 | 24.25 ± 0.37 | 23.07 ± 1.25 | 0.601 ± 0.010 | 0.603 ± 0.011 | 0.522 ± 0.037 | 470 |
| Iterative DNN-HIO [43] | 24.54 ± 0.34 | 24.54 ± 0.35 | **24.48 ± 1.12** | 0.692 ± 0.010 | 0.693 ± 0.011 | **0.665 ± 0.038** | 178 |
| Plug-and-Play PR | 21.67 ± 0.20 | 21.66 ± 0.21 | 22.23 ± 1.95 | 0.575 ± 0.007 | 0.575 ± 0.008 | 0.583 ± 0.040 | 202 |
| Plug-and-Play HIO | **25.26 ± 0.33** | **25.29 ± 0.36** | 24.27 ± 1.42 | **0.712 ± 0.012** | **0.714 ± 0.013** | 0.645 ± 0.031 | 248 |
| $\alpha = 4$ | Avg. PSNR (dB) | | | Avg. SSIM | | | Runtime |
| (Avg. SNR: 30.44 dB) | Overall | Natural | Other | Overall | Natural | Other | (sec.) |
| HIO | 19.01 ± 0.06 | 18.98 ± 0.05 | 20.09 ± 0.66 | 0.309 ± 0.002 | 0.307 ± 0.002 | 0.387 ± 0.017 | **162** |
| PrDeep [18] | 22.88 ± 0.22 | 22.90 ± 0.21 | 21.99 ± 1.33 | 0.539 ± 0.007 | 0.541 ± 0.007 | 0.494 ± 0.033 | 475 |
| Iterative DNN-HIO [43] | 23.46 ± 0.15 | 23.47 ± 0.14 | **23.02 ± 1.09** | 0.637 ± 0.006 | 0.637 ± 0.006 | **0.640 ± 0.036** | 179 |
| Plug-and-Play PR | 21.17 ± 0.12 | 21.17 ± 0.11 | 21.08 ± 1.97 | 0.550 ± 0.005 | 0.550 ± 0.005 | 0.564 ± 0.030 | 201 |
| Plug-and-Play HIO | **23.64 ± 0.24** | **23.66 ± 0.22** | 22.99 ± 1.33 | **0.644 ± 0.009** | **0.644 ± 0.009** | 0.609 ± 0.036 | 247 |

To evaluate the reconstruction quality visually, sample reconstructions for the natural images "Tiger" and "Bird" from the test dataset are also shown in Fig. 1. As seen from the images and zoomed-in regions, plug-and-play HIO does not introduce much artifacts and over-smoothing like the other methods, and thereby provides the best reconstruction based on both visual inspection and quantitative comparison.

Plug-and-Play HIO also provides robust performance for different noise levels. As seen in the table, its reconstruction quality surpasses all the other methods at different noise levels including $\alpha = 2, 4$, although the parameters of the method are optimized only for $\alpha = 3$.

Moreover, the developed plug-and-play HIO method is capable of successfully reconstructing other type of images than natural ones although the used denoiser is trained only with natural images. While PrDeep method also exploits a denoiser prior learned solely from natural images, our method performs better. For visual comparison, sample reconstructions for the image "Galaxy" are illustrated in Fig. 1. As seen, while PrDeep reconstruction has artifacts, plug-and-play HIO reconstruction is almost artifact-free.

Plug-and-Play HIO also significantly outperforms Plug-and-Play PR in terms of both PSNR and SSIM. Sample reconstructions are given in Fig. 2 to visually show the effectiveness of replacing the ER iterations with the more effective HIO iterations when space-domain constraints are available. Plug-and-Play PR suffers from various reconstruction artifacts and slow convergence similar to the ER algorithm, while Plug-and-Play HIO, which replaces the ER step in the developed PNP method with few HIO iterations, provides much better reconstructions both visually and quantitatively as expected. Here we note that allowing more iterations in PnP PR by optimizing its parameters could slightly improve its reconstruction quality, but since HIO was a better choice than ER in this setting with real-valuedness and non-negativity constraints, we did not explore this further.

Due to the complex convergence behavior of HIO-type algorithms, we provide convergence plots for the Plug-and-Play HIO method to demonstrate its numerical stability. For randomly chosen five test images, Fig. 3 shows how the normalized difference ($\|z_{k+1} - z_k\|_2 / \|z_k\|_2$) and PSNR change with the iterations. As seen, for all cases, the normalized difference between the current and previous estimate converges to zero as desired. The jumps in these plots correspond to the iterations where the used DNN is changed. The used denoiser consists of 25 different DNNs trained for different noise levels with standard deviations taken from the interval [1, 50] with a step size of 2. In each iteration $k$, we use the DNN trained for the noise standard deviation closest to the desired $\sigma_k$. This results in using the same DNN in subsequent iterations until the next change. When we switch to a different DNN due to decreasing $\sigma_k$, a larger difference occurs between the current and previous estimate, causing a jump in the convergence plots.

Because phase retrieval algorithms are generally sensitive to initialization, we also evaluate the robustness of the developed approaches to different initializations and image characteristics. Fig. 4 shows the PSNR and SSIM histograms of each method when $\alpha = 3$. These histograms are computed using reconstructions of 236 distinct test images obtained with 5 Monte Carlo runs, meaning that 5 different initializations are used for the reconstruction of each image. In these histograms, Plug and Play HIO has more reconstructions at higher PSNR ($> 30$ dB) and SSIM ($> 0.9$) zones than the other methods in addition to performing better in terms of average PSNR and SSIM. This suggests that PnP HIO is less sensitive to initialization and different image characteristics compared to the other methods.

As mentioned before, since the compared algorithms, prDeep and iterative HIO-DNN, were originally initialized with the HIO reconstruction obtained with a particular procedure proposed in [18], the same initialization is used for the developed methods to enable a fair comparison. Nevertheless, to further analyze sensitivity to initialization, we conduct an additional experiment for Plug-and-Play HIO using only a single random initialization (for $\alpha = 3$ case). For this analysis, the algorithm is run for different number of inner iterations $L$ and outer iterations $T$, namely for $\{L = 5, T = 200\}$, $\{L = 20, T = 200\}$, $\{L = 50, T = 200\}$, and $\{L = 50, T = 400\}$. The results are shown in Fig. 5 for the test image "Tiger". As seen, when both $L$ and $T$ are increased as

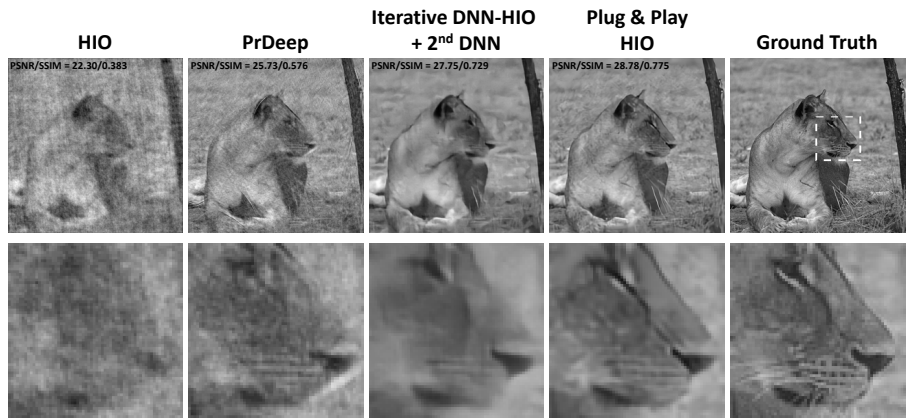

(a) Sample reconstructions for the natural image "Tiger".

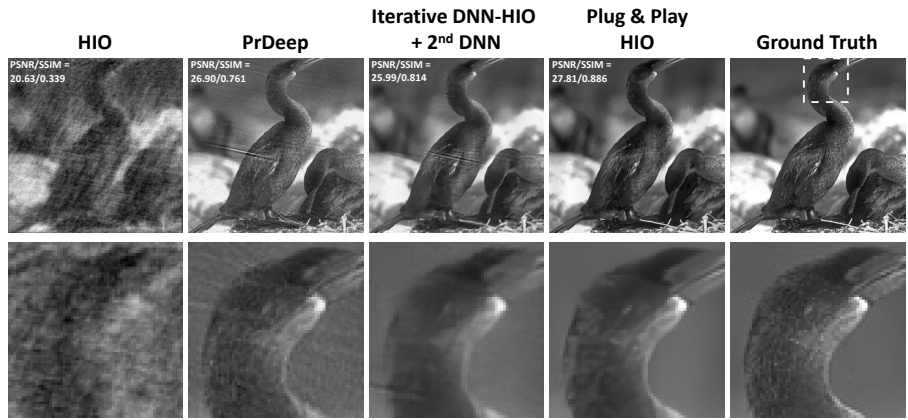

(b) Sample reconstructions for the natural image "Bird".

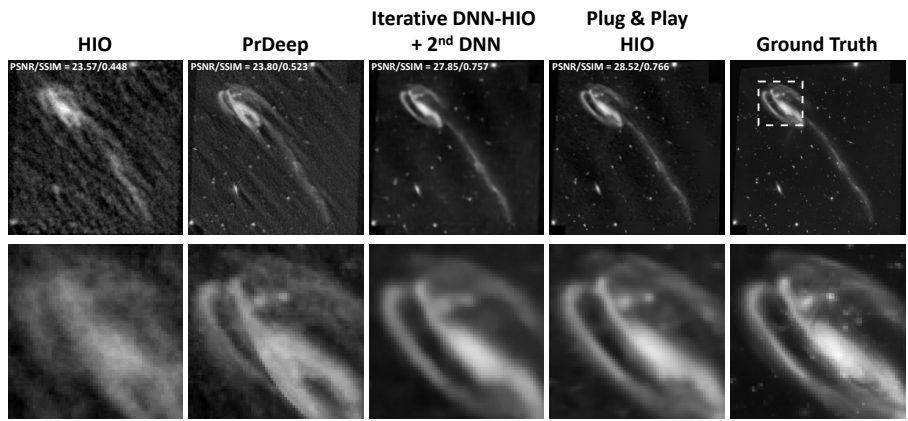

(c) Sample reconstructions for the image "Galaxy".

Figure 1. Reconstructions obtained with different methods for three test images and $\alpha=3$ case.

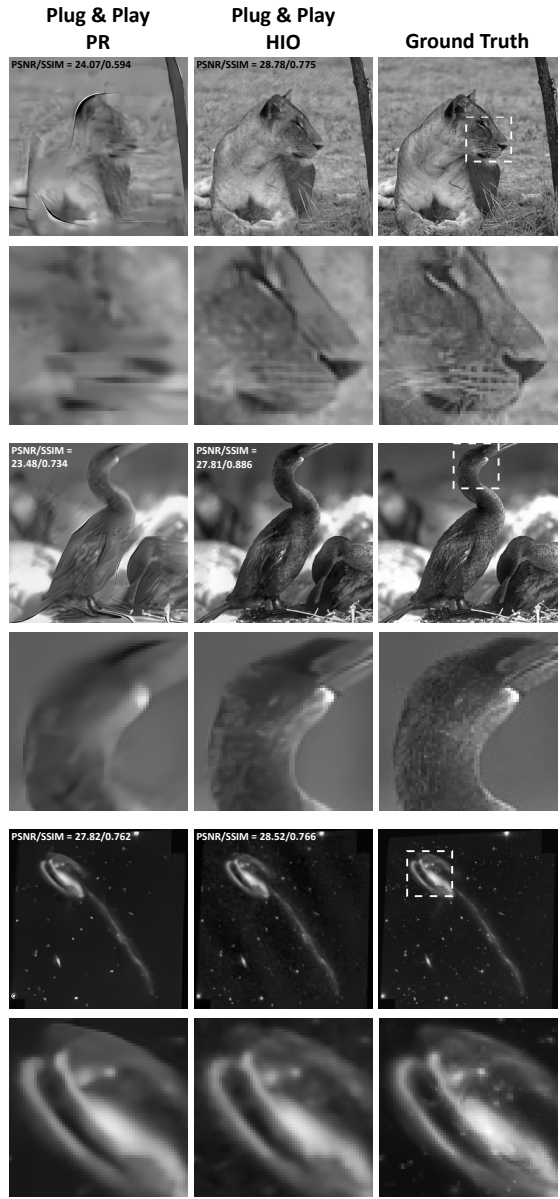

Figure 2. Comparison of PnP PR and PnP HIO

$L = 50$ and $T = 400$, a reconstruction with a quality similar to the earlier result in Fig. 1 has been obtained, which reaches to a PSNR of 25.94 dB. Hence the algorithm can be make more robust to initialization by choosing its parameters properly for the random initialization case.

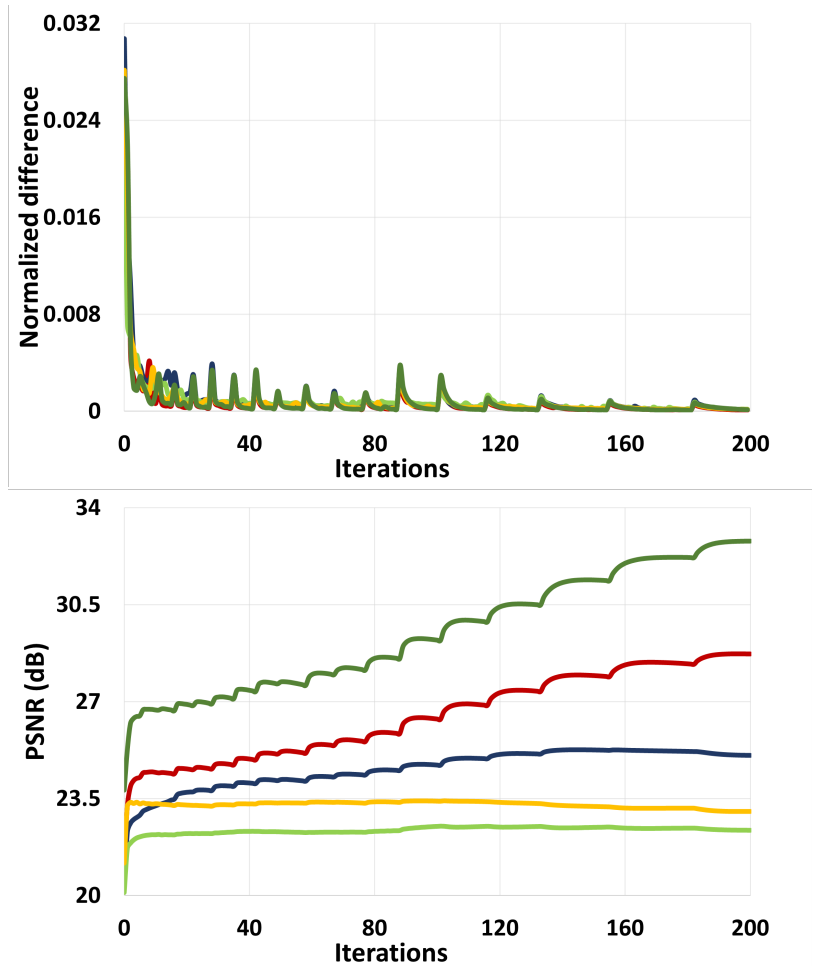

Figure 3. Normalized difference ($\|z_{k+1} - z_k\|_2 / \|z_k\|_2$) and PSNR versus iteration count for randomly chosen five test images to demonstrate the convergence of PnP HIO.

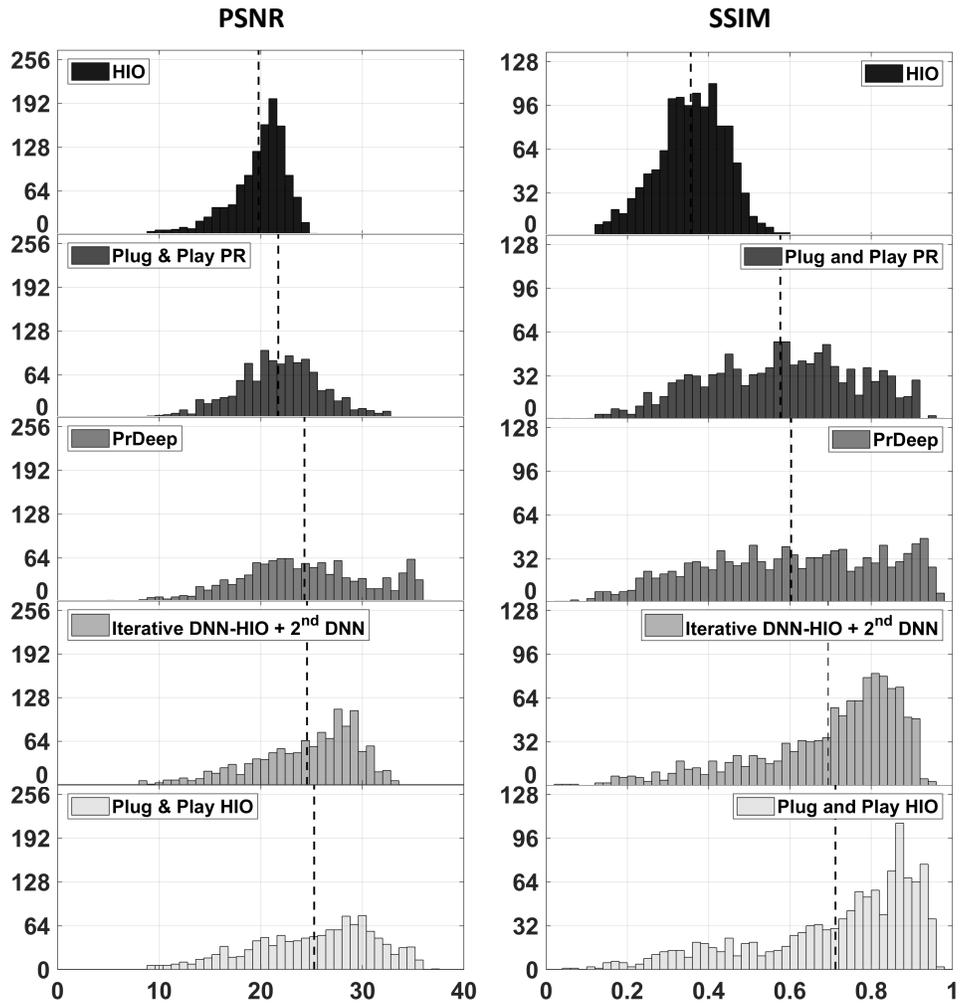

Figure 4. The histograms of PSNR (left column) and SSIM (right column) for the reconstructions obtained with different methods for all test images and 5 Monte Carlo runs ($\alpha = 3$ case). The average PSNR and SSIM values are shown with the vertical dashed lines.

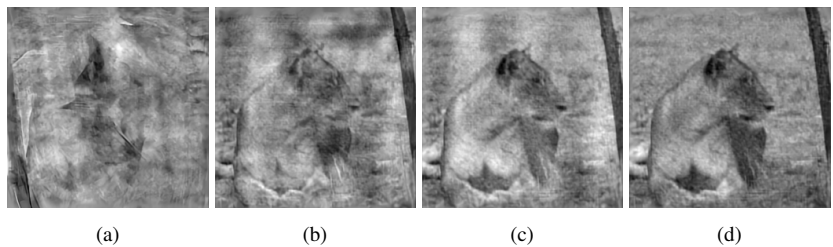

(a) (b) (c) (d)

Figure 5. Reconstructions of PnP HIO with a single random initialization for different number of inner iterations $L$ and outer iterations $T$: (a) $L = 5$, $T = 200$, (b) $L = 20$, $T = 200$, (c) $L = 50$, $T = 200$, (d) $L = 50$, $T = 400$.

## 5. Conclusions

In this paper, using HQS, we derived plug-and-play phase retrieval approaches, namely PnP PR and PnP HIO, which are iterative model-based methods that involve efficient update steps with learning-based prior. The methods perform efficient analytical update steps followed by learning based denoising. Analytical update steps consist of measurement and image update steps. In the measurement update step, convex combination of the measured intensity with the estimated measurement for the current guess is computed. Due to a varying weight parameter, the updated measurement is affected less from the actual measurement as the iterations proceed. In the image update step, the updated measurement is imposed on the current guess along with the available space domain constraints using projections (similar to the conventional ER/HIO methods). This intermediate result is then enhanced by a deep denoiser. By successively repeating these analytical update and denoising steps, final reconstruction is obtained. The effectiveness of the developed methods is demonstrated for Fourier phase retrieval problem in terms of reconstruction quality, computational efficiency, robustness to different initialization and noise levels using a large test dataset. The results demonstrate that Plug-and-Play HIO method not only provides state-of-the-art reconstruction performance, but also enables fast computation.

The plug-and-play phase retrieval approach derived in this paper (first presented in [2]) can be viewed as a regularized version of ER/HIO algorithms with a deep learning-based prior. As a result, the approach demonstrates the strengths of both the classical phase retrieval methods and deep learning-based regularization methods, where its computational efficiency and flexibility is inherited from the classical methods and its high reconstruction quality and robustness are thanks to the deep priors. Moreover, any denoiser algorithm and GS-type method can be easily exploited within the approach. In this work, the performance is demonstrated with a CNN-based denoiser, but using more recent deep denoisers such as diffusion models can help to improve the performance further.

Different than the learning-based direct inversion and unrolled methods, the presented approach has the advantage of applicability to different magnitude measurements without requiring any additional training. Hence the approach is flexible to be used with different measurements including Fourier magnitude, ptychography, and CDP measurements. In this paper, the performance is illustrated for the challenging case that involves Fourier magnitude measurements of real-valued images. Nevertheless, the approach is presented in general form and hence can be applied to other phase retrieval problems including the recovery of complex-valued images, which presents an exciting avenue for future work.

While the developed PnP approaches can work with any pre-trained denoiser, they still require supervised learning, meaning that a large training set is needed to optimize the weights of the denoiser network. Recently untrained DNNs are also employed as prior models for images. Such unsupervised learning-based approaches [59–61] do not require any labeled data for training. Although these approaches can require longer reconstruction time, they can offer self-calibrating capability, requiring only a rough prior knowledge of the imaging system. Due to the significance of these benefits in practice, future work could also explore exploiting untrained DNNs to regularize GS-type alternating projection methods.

This work demonstrates that, for phase retrieval, classical model-based methods can be substantially improved through the incorporation of deep priors via PnP regularization. We believe that the joint use of model-based methods with deep learning will play an important role in developing robust and efficient algorithms for nonlinear inverse problems.

**Funding.** Türkiye Bilimsel ve Teknolojik Araştırma Kurumu (120E505).

**Acknowledgment.** This study was funded by Scientific and Technological Research Council of Turkey (TUBITAK) under the Grant Number 120E505. Figen S. Oktem thanks TUBITAK for the support.

**Disclosures.** The authors declare no conflicts of interest related to this article.

**Data availability.** Data underlying the results presented in this paper are not publicly available at this time but may be obtained from authors upon reasonable request.